\documentclass[conference]{IEEEtran}
\IEEEoverridecommandlockouts
\usepackage{cite}
\usepackage{amsmath,amssymb,amsfonts}
\usepackage{algorithmic}
\usepackage{graphicx}
\usepackage{textcomp}
\usepackage{xcolor}
\usepackage{rotating}
\usepackage{hyperref}
\usepackage{listings}
\usepackage{caption} 
\usepackage{float}
\usepackage{subcaption}
\usepackage{booktabs}
\def\BibTeX{{\rm B\kern-.05em{\sc i\kern-.025em b}\kern-.08em
    T\kern-.1667em\lower.7ex\hbox{E}\kern-.125emX}}
\newcommand{\linebreakand}{%
  \end{@IEEEauthorhalign}
  \hfill\mbox{}\par
  \mbox{}\hfill\begin{@IEEEauthorhalign}
}

\begin{document}

\title{Quranic Audio Dataset: Crowdsourced and Labeled Recitation from Non-Arabic Speakers
}

\author{
\IEEEauthorblockN{1\textsuperscript{st} Raghad Salameh}
\IEEEauthorblockA{\textit{Innopolis University} \\
                 Republic of Tatarstan, Russia \\
                 r.salameh@innopolis.university\\
                  }
\and
\IEEEauthorblockN{2\textsuperscript{nd} Mohamad Al Mdfaa}
\IEEEauthorblockA{\textit{Skoltech University} \\
                 Moscow, Russia \\
                 mohamad.almdfaa@skoltech.ru\\
                  }
\hfill
\IEEEauthorblockN{4\textsuperscript{th} Manuel Mazzara}
\IEEEauthorblockA{\textit{Innopolis University} \\
                 Republic of Tatarstan, Russia \\
                 m.mazzara@innopolis.ru} 

\and
\IEEEauthorblockN{3\textsuperscript{rd} Nursultan Askarbekuly}
\IEEEauthorblockA{\textit{Innopolis University} \\
                 Republic of Tatarstan, Russia \\
                 n.askarbekuly@innopolis.university\\
                  } 
}

\maketitle

\begin{abstract}
This paper addresses the challenge of learning to recite the Quran for non-Arabic speakers. We explore the possibility of crowdsourcing a carefully annotated Quranic dataset, on top of which AI models can be built to simplify the learning process. In particular, we use the volunteer-based crowdsourcing genre and implement a crowdsourcing API to gather audio assets. We integrated the API into an existing mobile application called NamazApp to collect audio recitations. We developed a crowdsourcing platform called Quran Voice for annotating the gathered audio assets. As a result, we have collected around 7000 Quranic recitations from a pool of 1287 participants across more than 11 non-Arabic countries, and we have annotated 1166 recitations from the dataset in six categories. We have achieved a crowd accuracy of 0.77, an inter-rater agreement of 0.63 between the annotators, and 0.89 between the labels assigned by the algorithm and the expert judgments.\footnote{ For access to related resources including the Quranic Audio Dataset, please visit \href{https://quranic-audio-dataset.github.io/}{the paper webpage}.}
\end{abstract}


\section{Introduction}
The Quran is the Holy Scripture of Islam, and millions of Muslims are learning to read and recite it all over the world. Learning how to recite the Quran correctly in the original Arabic form is obligatory, however, many Muslims are of non-Arabic descent and do not speak the language. Normally, this problem is solved through learning with a qualified teacher, however, not everyone has access to a human instructor. AI technology can be used to partially cover this need, and simplify the process of learning the recitation of the Quran. In particular, speech recognition models can be used to detect mistakes in recitation and provide feedback on the proficiency of the learner \cite{Al-Ayyoub2018}, \cite{Al-Bakeri2017}, \cite{Bettayeb2021}. Training such models requires a large amount of labeled data, more specifically the recordings of recitations from many different people with common mistakes and the corresponding labeling.

In this research, we worked on validating the following hypothesis:
\begin{enumerate}
    \item A dataset of Quranic recitation audios can be crowdsourced from beginner learners via a recitation app.
    \item The collected dataset can be labeled through a dedicated crowdsourcing tool.
\end{enumerate}

Our study aimed to explore two key questions: Are beginner learners willing to share their voices when reciting the Quran, and are proficient reciters willing to participate in labeling audio recordings of recitations? To investigate these hypotheses, we followed a multi-step approach. First, we constructed a dataset through crowdsourcing, which involved integrating our project with the mobile application "NamazApp" to collect Quranic data. Next, we developed a crowdsourcing tool, dubbed "Quran Voice", to preprocess and label the collected data. Finally, we conducted manual quality control to validate the data, ensuring its readiness for use in training machine learning models.

The rest of the paper is structured as follows. We provide a brief overview of the existing work for Arabic linguistic resources and the Quran in Section \ref{sec:lr}. We then discuss the process of collecting the data and the annotation collection procedures in Section \ref{sec:methodology}. We present detailed information about the collected dataset in Section \ref{sec:result}. Finally, we conclude our work in Section \ref{sec:con}.

\section{Related work}
\label{sec:lr}
Crowdsourcing is the practice of assigning a computationally challenging task to a large, typically online group of people. It is a good way to break down a manual, large, time-consuming task into smaller, more manageable tasks to be completed by distributed workers \cite{Estellés2015}. 

Whenever AI research is conducted, datasets are considered a crucial component. The creation of a high-quality dataset requires a lot of effort and time.

There are three types of crowdsourcing paradigms for dataset creation: mechanized labor (paid-for) \cite{b2} \cite{b3}, where employees receive cash compensation; games with a purpose (GWAP) \cite{b9} \cite{b11}, where the task is disguised as a game; and altruistic work (volunteer-based)  \cite{b5} \cite{b6}, which relies on goodwill. Any of the three prominent crowdsourcing paradigms can be employed.

\subsection{Crowdsourcing for Arabic linguistic resources}
Many Arabic speech researchers used Crowdsourcing for both corpus annotation text-based \cite{b2} \cite{b3} \cite{b4}, speech-based \cite{b5} \cite{b6} \cite{b7} \cite{b8} \cite{b9} \cite{b10}, and Arabic digitization with diacritics \cite{b11}.

In \cite{b2}, the authors aim to construct \textbf{Comment Dataset for Offensive Language Detection}. As a result, they annotated 4000 comments with an accuracy of 94\% using Amazon Mechanical Turk.

In \cite{b3}, the authors outline a two-step methodology for \textbf{annotating Arabic targets of opinions} using Amazon Mechanical Turk. First, they asked annotators to identify candidate targets “entities” in a given text. Then, they asked annotators to identify the opinion polarity (positive, negative, or neutral) expressed about a specific entity.

In \cite{b4}, the authors were running experiments to see if \textbf{Crowdsourcing can be used for Effective Annotation of Arabic}. They contrasted two methods for linguistically annotating an Arabic corpus: part-of-speech (POS) tagging and grammatical case-ending. The overall results were 50.07\% accuracy for grammatical case endings and 63.91\% for POS tagging. The study demonstrates that crowdsourcing for Arabic linguistic annotation is ineffective for both objectives since it calls for experienced annotators.

In \cite{b5}, the authors aim to create \textbf{Speech Corpus for Algerian Arabic Dialectal} called Kalam’DZ which includes the 8 major Algerian Arabic sub-dialects with 4881 speakers and more than 104.4 hours. Most annotations are made manually by assigning for each utterance the spoken dialect and validating the speaker's gender.

In \cite{b6}, the authors aim to prove that using the altruistic (volunteer-based) approach is an effective approach by validating 10\% of dialect annotation conducted for Kalam’DZ \cite{b5} which equals 10 hours and 1012 tasks. The contributors are asked to answer a question about a given audio with one of the three possible responses (Yes, No, Unknown). The crowd annotation accuracy was about 81.1\%.

In \cite{b7}, the authors aim to build \textbf{ Massively-Multilingual Speech Corpus}. For \textbf{Data Collection} They collected 2,500 hours of audio from over 50,000 participants. The recordings are verified by contributors as correct or incorrect.

In \cite{b8}, the authors aim to gather large \textbf{speech corpora for Egyptian dialectal Arabic}. Using a designed Game With a Purpose, they collect transcriptions for 120 audio files with 1121 Arabic orthographic transcriptions and 1121 transcriptions in the Arabic Chat Alphabet.

In \cite{b10}, \textbf{A multi-dialectal speech corpus of DA}, produced by the authors from high-quality broadcast, debate, and discussion programs from Al Jazeera, incorporates both scripted and spontaneous speech. They obtained dialect labels for 57 hours of Egyptian, Levantine, Gulf, and North African DA. They automatically labeled an additional 94 hours by using speaker linking to recognize utterances made by the same speaker. 

In \cite{b9}, the authors present \textbf{a GWAP for crowdsourcing classifications of several dialects of Arabic in multi-dialectal audio}. Players in Lahajet choose a character that represents the accent after listening to brief audio snippets. 

In \cite{b11}, the authors present \textbf{tashkeelWAP: A GWAP For Digitizing Arabic Diacritics}. The participants need to validate images along with their corresponding diacritic-less digitization.

\subsection{Crowdsourcing for Quran}

Regarding Quran, there are extremely few public speech datasets for the Quran available. we can mention the following:

In \cite{b12}, the authors aim to build \textbf{dataset of crowd\-sourced Quranic recitation}. They collected 50,000 verses. About 150 verses were manually annotated. The entire dataset has also been automatically evaluated by using Google Speech-to-text to first transcript the recording, then using Iqra (a Quran search engine) to search for the verse using the transcription. If the result returned by Iqra matches the recorded ayah, they mark it as correct.

In \cite{b13}, the authors build a \textbf{dataset QDAT} from the recitation of one verse from the Quran. They gathered 1500 audio files having both correct and incorrect recitations. Experts manually annotate the audio files to demonstrate the accuracy of reciting the Quran with Tajweed while following three rules of recitation: Al Mad, Ghunnah, and Ikhfaa.

In \cite{b18}, the authors present a \textbf{database of Quranic recitations} for automatic speech processing based on tajweed correction. This database contains audio recordings from Surah AlFatihah where each audio recitation has some purposefully inserted tajweed (rules of recitations) errors. Experts identify 54 errors that reciters of Surah Al-Fatihah may make and collect samples for each such error case. 17 volunteers participated, Arabic and non-Arabic speakers. Each audio file is labeled for the case it covers, along with information on the reciters.

As can be seen, there are not enough public Quranic datasets available. This leads to each research project independently developing its own database to facilitate experimentation, resulting in challenges when comparing findings. Moreover, crowdsourcing has not been used in Quran annotation. Our research aims to fill these gaps and examine the willingness of beginner learners to share their recitations and proficient reciters to label the gathered data.

\section{Initial Methodology Plan}
\label{sec:methodology}
To construct a high-quality Quranic recitation dataset, we have followed the crowdsourcing engineering process defined by \cite{b14}. It suggested breaking down the process into four main stages: Project Definition, Data Preparation, Project Execution, and Data Evaluation and Aggregation.

\subsection{Project Definition}
The first step is to choose the appropriate crowdsourcing genre. In this project, we choose the altruistic work crowdsourcing genre.
Secondly, the chosen NLP problem needs to be decomposed into a set of simple crowdsourcing tasks. We can define the tasks for our project as the following:

\begin{enumerate}
    \item Beginner learners are asked to record their voices while reciting a specific verse of the Quran.
    \item Proficient reciters are asked to listen to audio and provide the correct classification.
\end{enumerate}

\subsection{Data Preparation}
Besides collecting and preparing the data, we must design the crowdsourcing user interfaces.
\subsubsection{The first crowdsourcing task}
For the first crowdsourcing task, we built a Quran player responsible for collecting Quranic recitation coupled with, the chapter and verse being recited, and demographic background information about the user (id, age, gender, country, platform, Qiraah). The player contains short chapters of the Quran and gives the users the ability to:
\begin{itemize}
    \item See and listen to the desired verse (Aya) being recited by a Qari (professional reciter).
    \item Record their voice.
    \item Compare their recording with the one from the Qari.
    \item Upload their recorded voice if they want to share it and receive feedback. 
\end{itemize}

\begin{figure*}[thpb]
    \centering
    \begin{subfigure}{0.39\textwidth}
        \centering
        \includegraphics[width=0.6\linewidth]{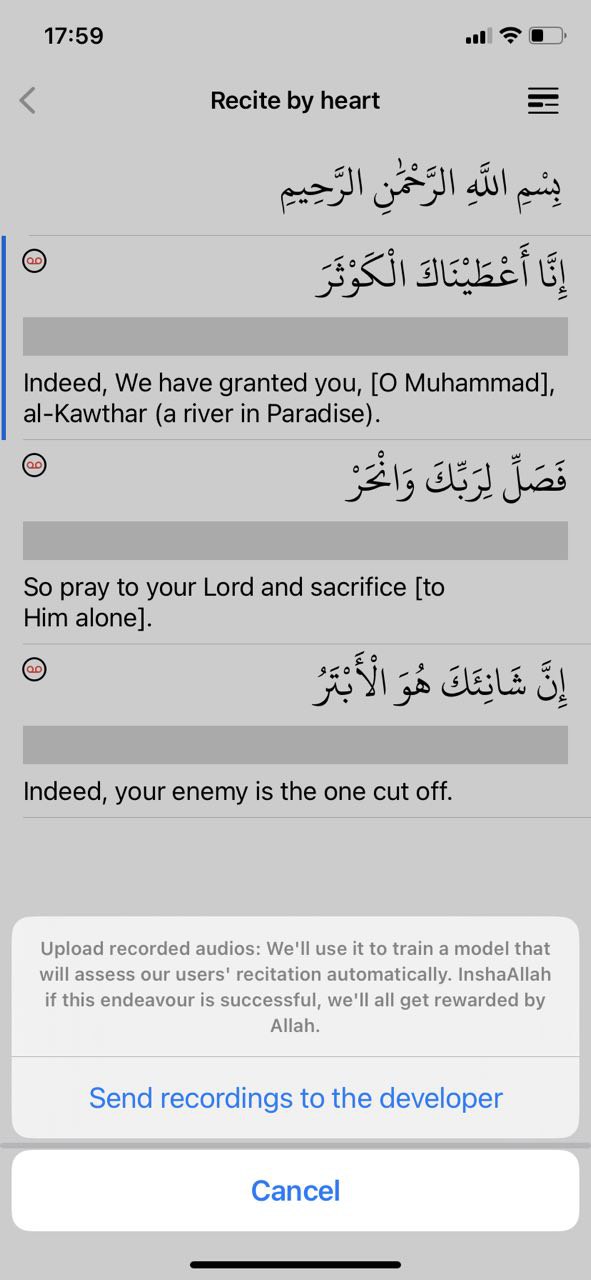}
        \caption{Quran player - Namaz App}
        \label{fig:share recitation}
    \end{subfigure}%
    \begin{subfigure}{0.58\textwidth}
        \centering
        \includegraphics[width=0.9\linewidth]{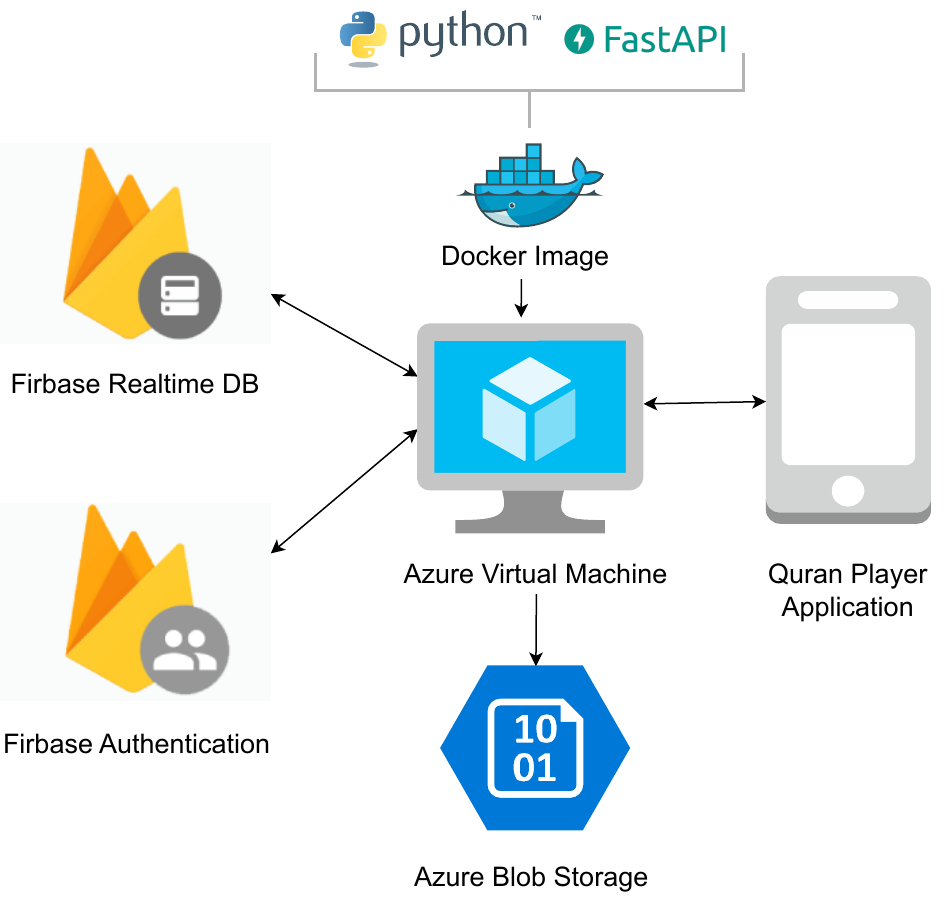}
        \caption{Quran player backend Architecture}
        \label{fig:Quran player backend}
    \end{subfigure}
    \caption{Quran player Architecture for the first crowdsourcing task}
    \label{fig:Quran player Architecture}
\end{figure*}

Importantly, the app provides value for beginner learners since it will help them to learn how to recite Quran correctly when receiving feedback. Additionally, they can choose to share the recordings of their recitations as shown in Figure \ref{fig:share recitation}. The Architecture for the Quran player is shown in Figure \ref{fig:Quran player backend}.

\begin{figure}[H]
    \centering
    \includegraphics[scale=0.4]{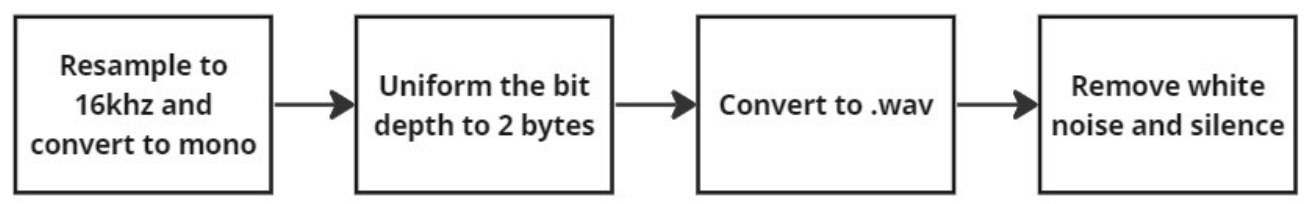}
    \caption{Audio standardization pipeline}
    \label{fig:Audio standardization pipeline}
\end{figure}

Before storing the audio of the recitation on the server, we make it pass through our audio standardization pipeline to ensure consistency and compatibility across audio files. the pipeline stages are shown in Figure \ref{fig:Audio standardization pipeline}.
We employed the Pydub Python library to standardize all audio files by converting them to the WAV format with a mono channel, 16kHz frequency, and 2 bytes bit-depth in addition to removing silence segments. Moreover, we utilized the Noisereduce library \cite{b19} to effectively  eliminate background noise from audio recordings. 

The Quran player was implemented as an importable package that can be integrated easily by other apps. Currently, this player is already imported into \href{https://apps.apple.com/us/app/namaz-app-learn-salah-prayer/id1447056625}{Namaz App} which was published on App Store.

\subsubsection{The second crowdsourcing task}
\begin{figure*}[thpb]
    \centering
    \includegraphics[width=\textwidth]{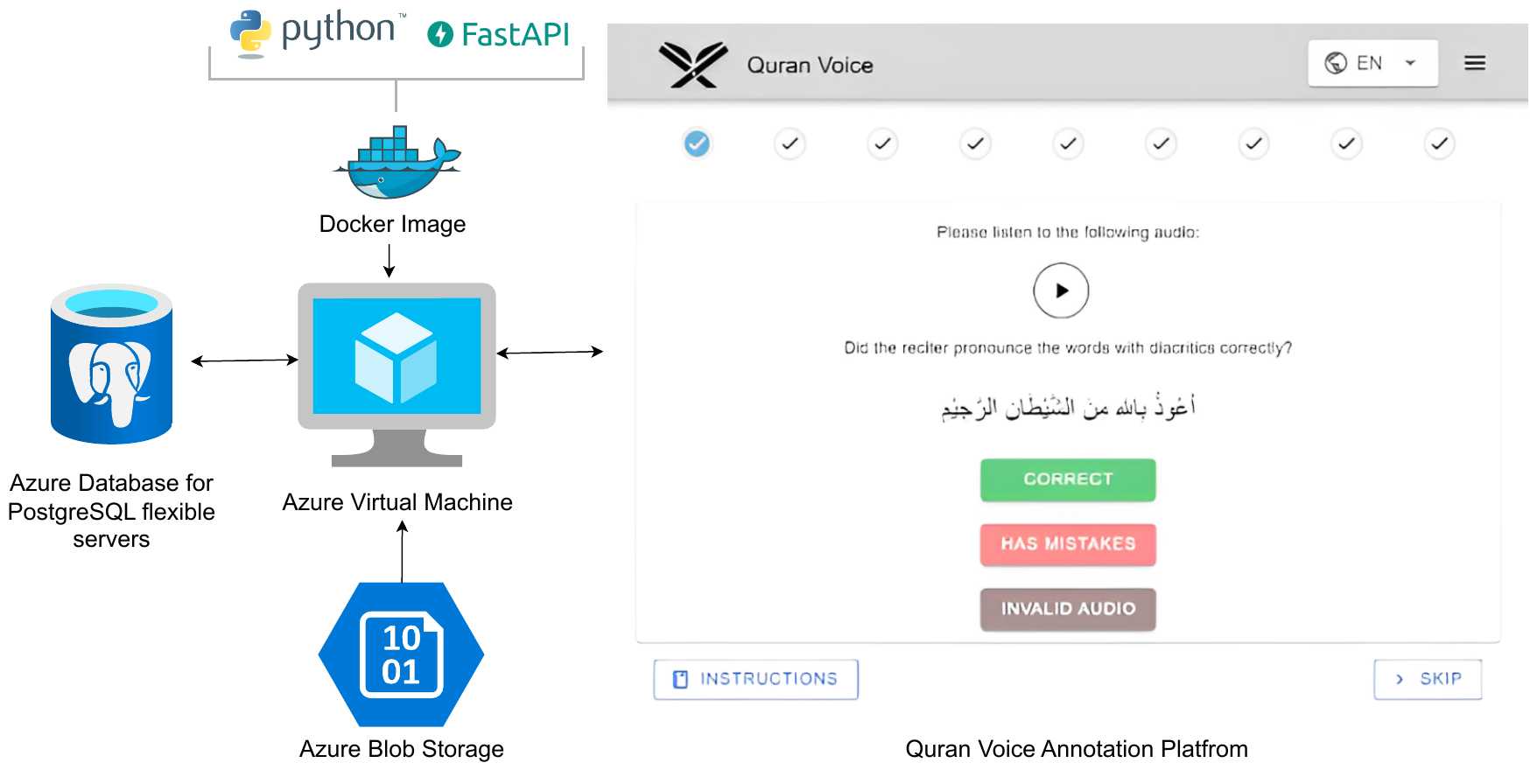}
     \caption{Quran Voice Architecture for the second crowdsourcing task}
    \label{fig:Quran Voice Architecture}
\end{figure*}

For the second crowdsourcing task, we sought to cover words and diacritics mistakes on the audio data we collected in the first crowdsourcing task. For this purpose, we implemented Quran voice web platform as shown in Figure \ref{fig:Quran Voice Architecture}. The website is available in three languages: Arabic, English, and Russian.

In this task, referred to as \textbf{Validate Verse Correctness}, proficient reciters can help us validate if the verse has been read correctly by validating the correctness of letters and diacritics. The participant were given an audio of a specific verse to listen to it and label it with one of the following choices:
\begin{itemize}
    \item Correct: when the pronunciation is correct with the diacritics, regardless of the rules of Tajweed.
    \item Has mistakes: The answer is incorrect when the pronunciation is wrong with the diacritics, regardless of the rules of Tajweed.
    \item Invalid Audio: When there is a problem with the audio.
    \begin{itemize}
        \item Empty / Not related: When the content of the audio clip is incomprehensible or contains words that have nothing to do with the Quran or empty, this choice should be selected.
        \item Different Verse: when the audio clip contains words related to the Quran, but not the given verse, this choice should be selected. 
        \item Multiple Verses: When a reciter reads several verses, this choice should be selected regardless of whether the reading is correct or not.
        \item Incomplete Verse: When the reciter reads the verse without completing it, or the verse is not completed for some reason, this choice should be selected.
    \end{itemize} 
\end{itemize}

\textbf{An annotation guideline} was prepared for this task with real samples from the dataset. In addition to written instructions, video instructions were also provided for those who prefer visual aids.

\begin{figure}[!ht]
    \centering
    \includegraphics[scale=0.12]{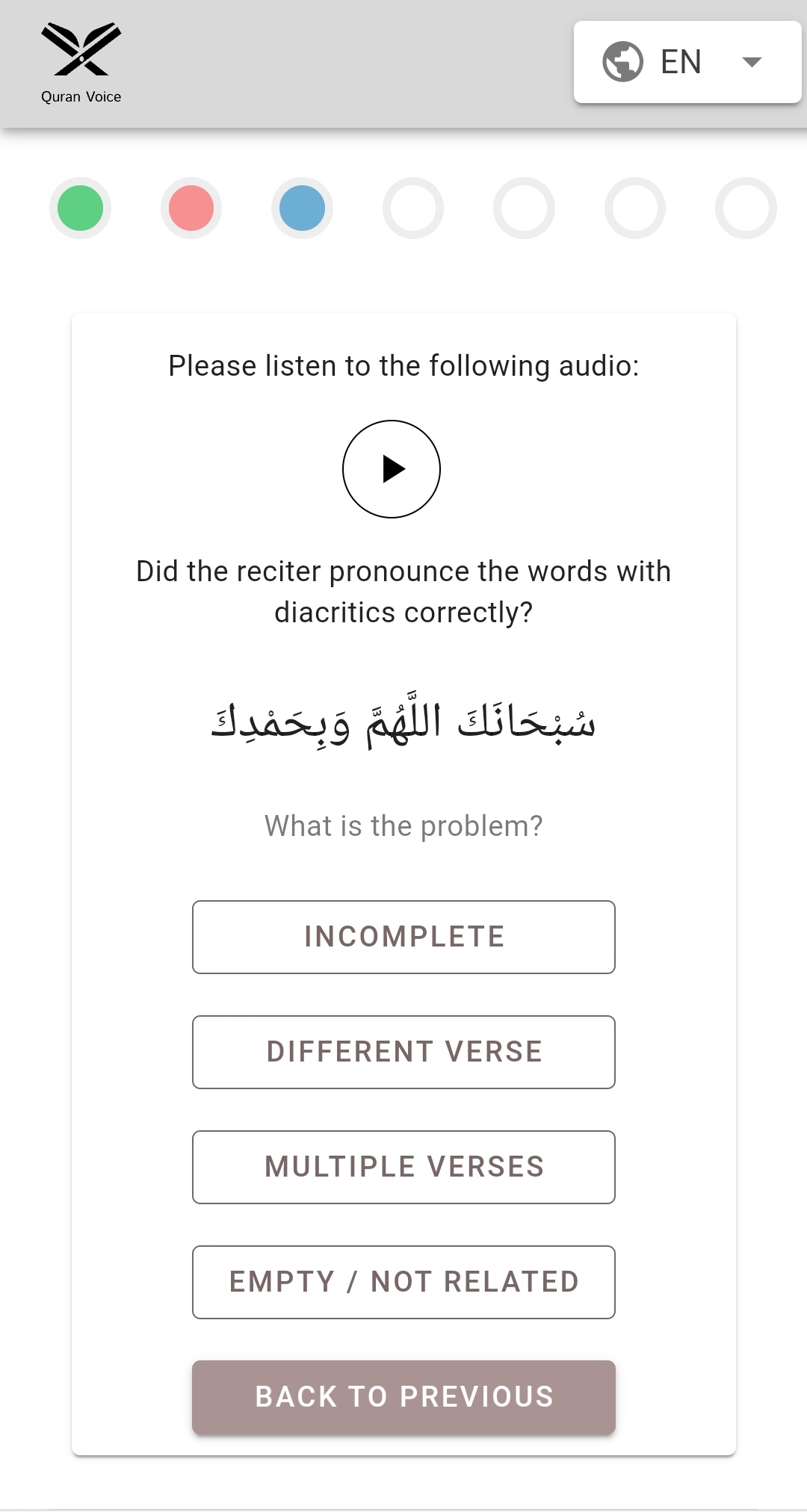}
    \captionsetup{justification=centering}
    \caption{Training session for validate Verse Correctness task on Quran Voice}
    \label{fig:training session}
\end{figure}

Before the participants access the real tasks, they have to pass the training session as shown in Figure \ref{fig:training session}. A set of 62 \textbf{gold standards tasks} were collected and annotated by experts for this purpose. The training session has 8 questions: 2 labeled as Correct, 2 have mistakes, and 1 for each remaining labels.
The participant passes the training session if they score at least 0.6  \textbf{Matthews Correlation Coefficient}. The participant has 5 attempts to pass the training session, otherwise, they will be excluded.

\subsection{Project Execution}
\label{sec:project-ex}
There are decisions to be taken, such as whether the complete dataset should be annotated more than once to enable a reconciliation and verification stage (better quality, but greater costs), or whether it is sufficient to have only two or three annotators per task as long as they can agree. Some of the studies mentioned in Section \ref{sec:lr} used 3 judgments and others used 5 judgments. For our study, we started by taking 3 judgments. The details are described in Section \ref{sec:agg}.

To ensure quality annotations, we implement a rigorous \textbf{quality control} process. In addition to the training session that participants should pass to access real tasks, we use control tasks to monitor their performance during the annotation process. The initial set of control tasks is manually annotated by experts, and we convert real tasks with full inter-annotator agreement to control tasks. This helps us adjust participants' scores and maintain high-quality annotations.

To encourage participation in the voluntary-based crowdsourcing approach, advertising efforts were made through social media. A dedicated Facebook page, called {\href{https://www.facebook.com/profile.php?id=100091268018169}{"Quran Voice"}}, was created to publish promotional posts for the website continuously.

\begin{figure}[!ht]
    \centering
    \includegraphics[scale=0.12]{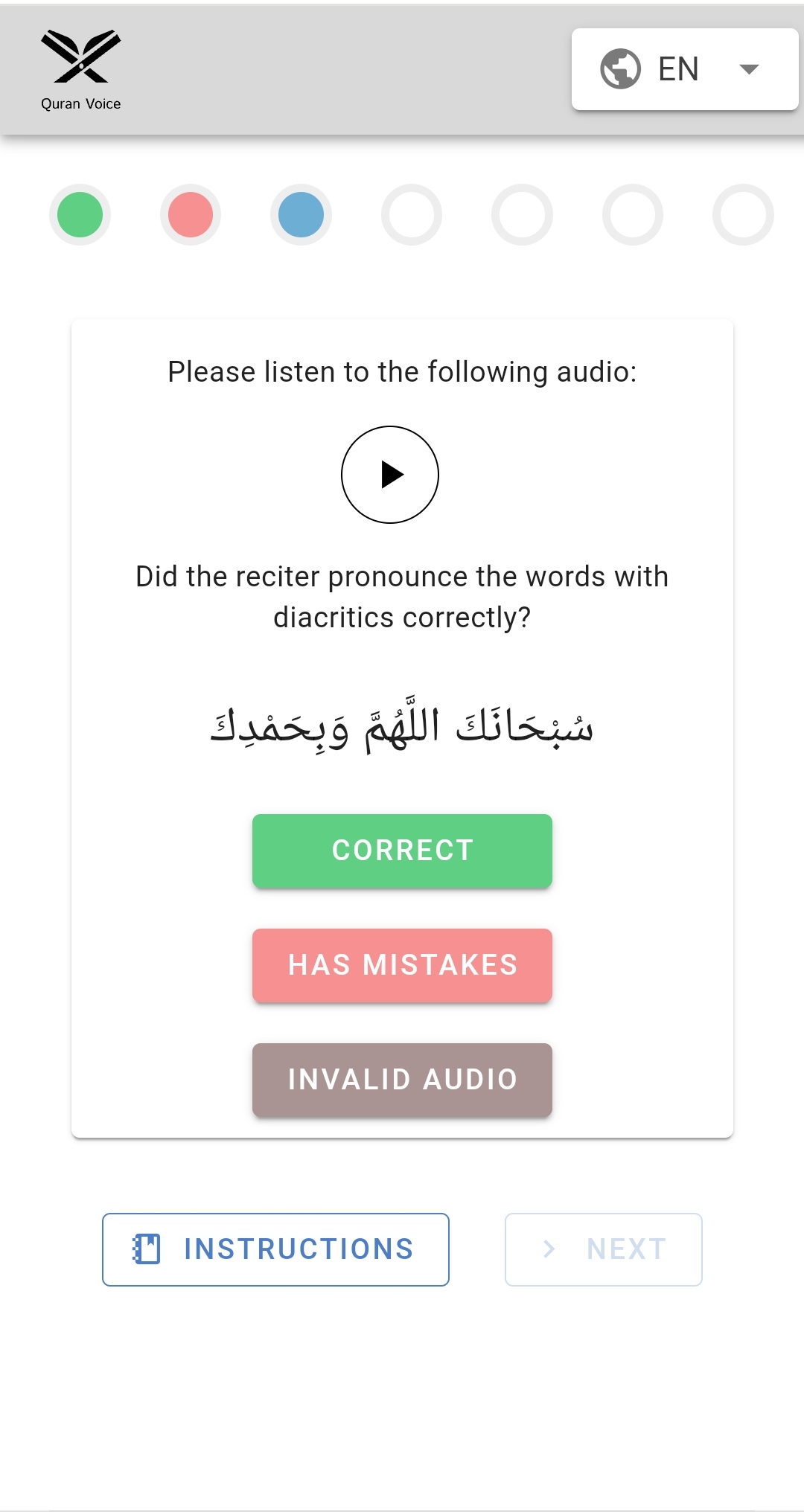}
\end{figure}

\subsection{Data Evaluation and Aggregation}
\label{sec:agg}
The difficulty in this stage is in analyzing and combining the contributions from various contributors into a comprehensive linguistic resource, and in evaluating the resulting overall quality.

\begin{figure}[!ht]
    \centering
    \includegraphics[scale=0.65]{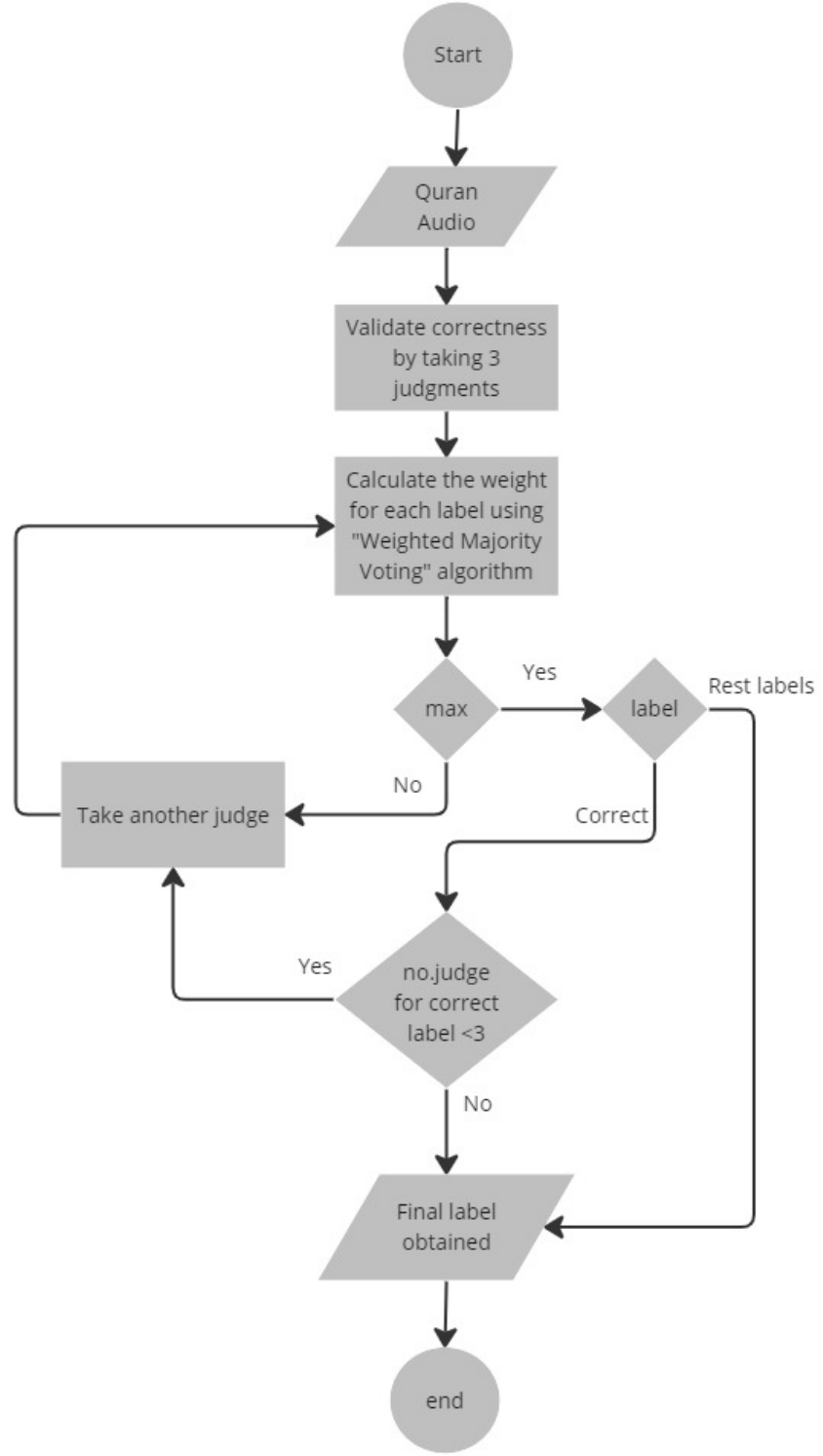 }
    \caption{Annotation Aggregation for Validate Verse Correctness task}
    \label{fig:Validate correctness}
\end{figure}

To aggregate the final annotations for the Validate Recitation task, we employ the \textbf{Weighted Majority Voting} algorithm, which is suitable for multi-class classification tasks. This approach is particularly useful in our study because it considers the performance score of participants as a weight for their answers. The aggregation process and the number of judges involved are illustrated in figure \ref{fig:Validate correctness}.

For data evaluation, We took 10\% of the data annotated by the crowd and let it be manually annotated by an expert.

\section{Results and Discussion}
\label{sec:result}
\subsection{ Overview of the Collected Datasets }
\begin{figure*}[thpb]
    \centering
    \begin{subfigure}{0.48\textwidth}
        \centering
        \includegraphics[width=\linewidth]{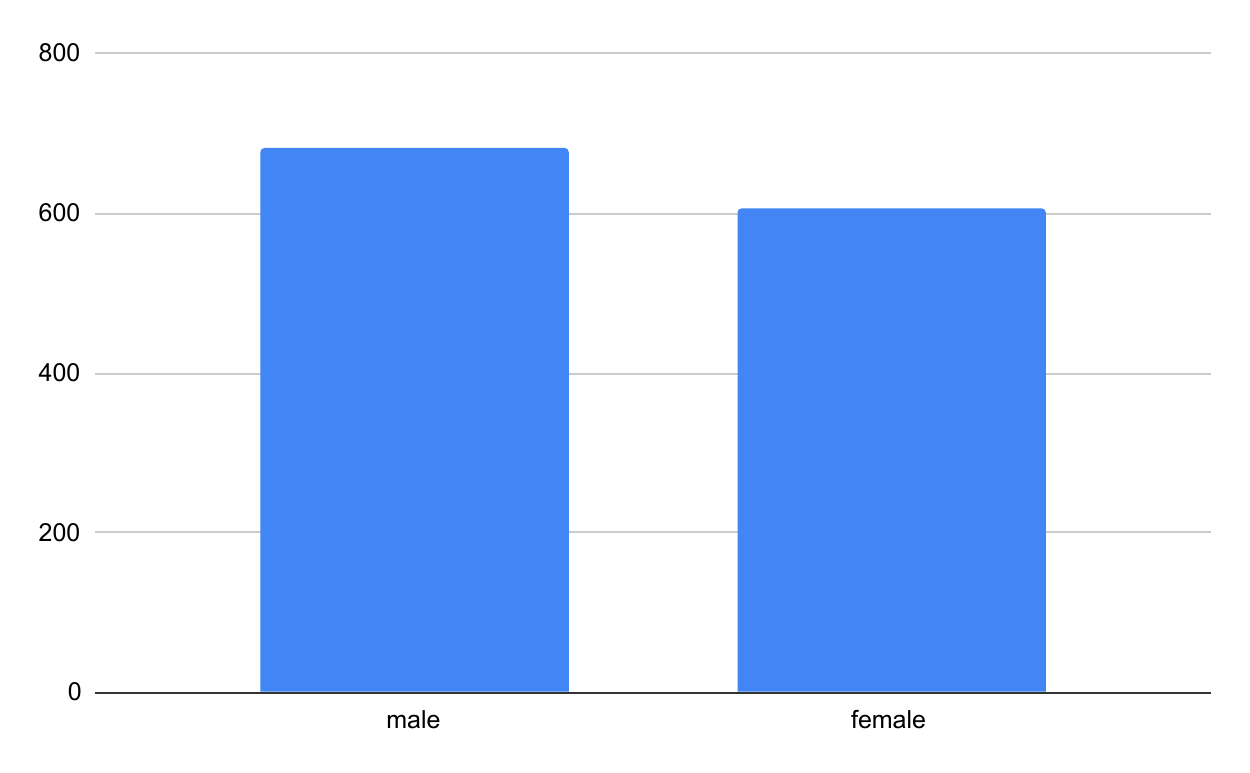}
        \caption{Gender Distribution}
        \label{fig:Gender Distribution}
    \end{subfigure}%
    \begin{subfigure}{0.48\textwidth}
        \centering
        \includegraphics[width=\linewidth]{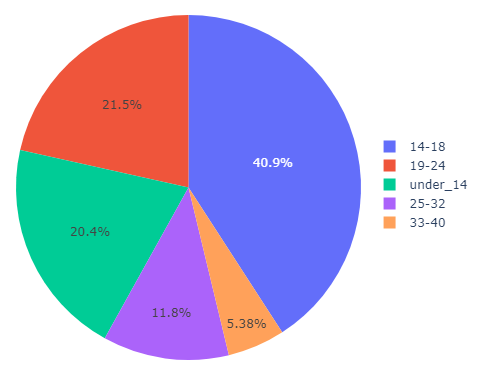}
        \caption{Reported Age Distribution}
        \label{fig:Age Distribution}
    \end{subfigure}
    \begin{subfigure}{0.48\textwidth}
        \centering
        \includegraphics[width=\linewidth]{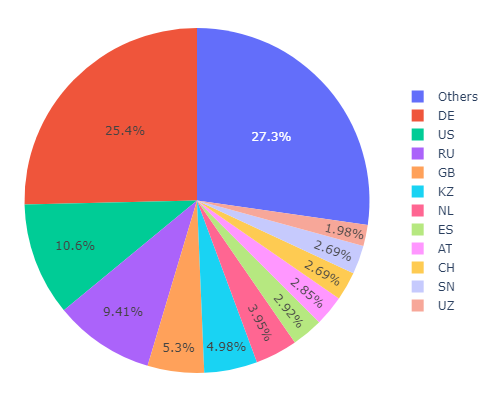}
        \caption{Global Distribution}
        \label{fig:Global distribution}
    \end{subfigure}%
    \begin{subfigure}{0.48\textwidth}
        \centering
        \includegraphics[width=\linewidth]{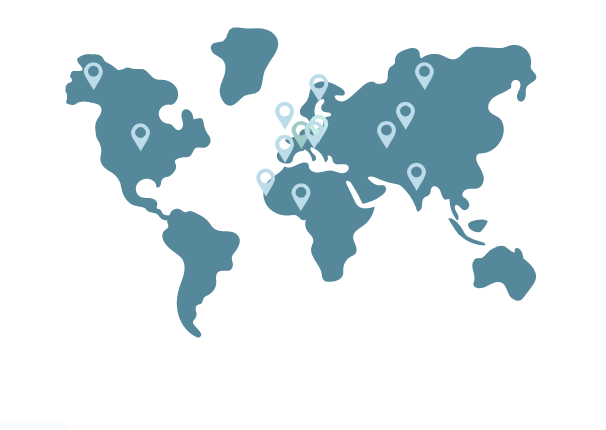}
        \caption{Global Distribution Map}
        \label{fig:Global distribution map}
    \end{subfigure}
    \caption{Demographic Profile of Participants in the Quranic Audio Dataset}
    \label{fig:profile}
\end{figure*}

In the first task, a pool of 1287 participants recorded their voices while reciting specific verses. As a result, we've recorded around 7000 verses, totaling 11.5 hours. Figure \ref{fig:profile} illustrates demographic information about the participants.

Among the participants, 47.1\% identified as female, and 52.9\% identified as male, as illustrated in Figure \ref{fig:Gender Distribution}. Out of the 1287 participants, only 93 provided their ages, with the distribution depicted in Figure \ref{fig:Age Distribution}. As shown in Figures \ref{fig:Global distribution}-\ref{fig:Global distribution map}, The majority of reciters are from 11 countries around the world, namely Germany, United States, Russia, Kazakhstan, United Kingdom, Netherlands, Spain, Austria, Switzerland, Senegal and Uzbekistan.

We collected short verses and Duas that Muslims use in prayer and daily life. The summary found in Table \ref{tab:audio-summary}.

\begin{table}[h]
\caption{Current audio dataset statistics}
\label{tab:audio-summary}
\begin{tabular*}{\hsize}{@{\extracolsep{\fill}}llllll@{}}
\toprule
Surah & Audios & Females & Males & Unknown & Duration \\ &&&&&(min)\\
\midrule
Al-Faatihah & 2934 & 1375 & 1119 & 440 & 294 \\
Al-NABAA & 30 & 0 & 30 & 0 & 3 \\
Al-Qadr & 132 & 40 & 92 & 0 & 16 \\
Al-Asr & 182 & 89 & 52 & 41 & 20 \\
Al-Humazah & 108 & 54 & 40 & 9 & 11 \\
Al-Fil & 63 & 34 & 29 & 0 & 6 \\
Quraish & 38 & 19 & 19 & 0 & 4 \\
Al-Maaoon & 57 & 31 & 26 & 0 & 4 \\
Al-Kauthar & 183 & 67 & 81 & 35 & 19 \\
Al-Kafiroon & 309 & 154 & 79 & 76 & 29 \\
An-Nasr & 97 & 50 & 35 & 12 & 11 \\
Al-Masad & 112 & 40 & 45 & 27 & 12 \\
Al-Ikhlas & 561 & 240 & 218 & 103 & 42 \\
Al-Falaq & 278 & 110 & 110 & 58 & 28 \\
An-Nas & 557 & 223 & 255 & 79 & 52 \\
Ayat Ramadan & 27 & 5 & 20 & 2 & 3 \\
Ayat al-Kursi & 350 & 124 & 169 & 57 & 40 \\
At-Tahiyyat & 182 & 102 & 74 & 6 & 20 \\
Subhanaka & 91 & 58 & 30 & 3 & 8 \\
Salawat & 84 & 34 & 40 & 10 & 9 \\
Dua Qunoot & 31 & 3 & 28 & 0 & 2 \\
Adhkar after prayer & 70 & 8 & 62 & 0 & 6 \\
Dua for Protection & 30 & 17 & 10 & 3 & 3 \\
Adhan & 322 & 54 & 268 & 0 & 34 \\
Iqamah Prayer & 66 & 18 & 48 & 0 & 5 \\
Dua from the Quran & 6 & 0 & 6 & 0 & 0 \\
\bottomrule
\end{tabular*}
\end{table}

\subsection{Overview of the Annotated Datasets}
\label{subsec-annoatated-dataset}
In the second task, We had 322 participants registered on Quran Voice platform, 148 attempted to do the training session, and 71 of them passed the training session and were ready to participate in the annotating process. 

Currently, we have collected 4117 annotations of 1427 recitations made by 62 unique participants, and out of 7000 recitations, we got labels for 1166 recitations, while the rest 261 recitations were waiting for additional judges to get their labels. The number of occurrences for each label is shown in Table \ref{tab:lables-summary}.

\begin{table}[h]
\caption{The number of occurrences for each category}
\label{tab:lables-summary}
\begin{tabular*}{\linewidth}{@{\extracolsep{\fill}}lll@{}}
\toprule
Label & Frequency & Percentage\\
\midrule
Correct & 396 & 33.96\% \\
Has mistake & 476 & 40.82\% \\
Incomplete Verse & 22 & 1.89\% \\
Different Verse & 53 & 4.55\% \\
Multiple Verses & 78 & 6.69\% \\
Empty / Not related & 141 & 12.09\% \\
\bottomrule
\end{tabular*}
\end{table}

\subsection{Crowd Accuracy}
To calculate how well the participants performs while solving the task we used the Matthews correlation coefficient (MCC) metric since it is a more reliable statistical rate that produces a high score only if the prediction obtained good results in all of the four confusion matrix categories (true positives, false negatives, true negatives, and false positives).

We have mentioned earlier in \ref{sec:project-ex} that a user is allowed to participate in solving the task if they pass the entrance exam, which consists of 8 questions. To pass the entrance exam, they should score $MCC\geq0.6$. This number was picked as a threshold after experiments for the number of mistakes that annotators can make in the different classes.
A participant will not pass if they make more than one mistake in labeling Correct or Has Mistakes tasks or make more than two mistakes in labeling the rest of the labels. Based on the data presented in the given table \ref{tab:num-attempts}, we can conclude that allowing users to make three attempts was a suitable number.


\begin{table}[h]
\caption{The number of attempts for passing the test} 
\label{tab:num-attempts}
\begin{tabular*}{\linewidth}{@{\extracolsep{\fill}}lll@{}}
\toprule
Number of attempts & Number of all users & Number of passed \\ && users\\
\midrule
1 & 111 & 48 \\
2 & 24 & 18\\
3 & 13 & 5\\
\bottomrule
\end{tabular*}
\end{table}

		
By computing the average Matthews Correlation Coefficient (MCC), Accuracy, and F1 score of individual annotators, the overall quality of the participants was evaluated. The analysis yielded an estimated overall quality with an AVG\_MCC of 0.68, AVG\_Accuracy of 0.77, and AVG\_F1score of 0.74. 

\subsection{Inter-Rater Agreement}
To assess Inter-Rater agreement, we used Krippendorff's alpha \cite{b20} to analyze 958 labeled tasks with three judges each. The resulting value of 0.63 indicates Substantial agreement among the judges. Out of the 958 tasks, 533 were in total agreement, 42 were in total disagreement, and 383 were in partial agreement. After examining the tasks with total disagreement, we observed that some users did not understand the instructions clearly, leading to confusion between "Multiple Verses" and "Different Verse", "Correct" and "Multiple Verses", "Has mistake" and " Incomplete Verse", and "Has mistake" and "Empty / Not related".
We have observed also that certain audios are not clear or combined between multiple labels, which highlights the need for a "Not Clear" label to address these issues.

To assess the algorithm's performance in label selection, a random sample of $50$ tasks from the annotated dataset was selected and labeled by experts. The evaluation yielded impressive results, with an accuracy of $0.94$, an $MCC$ of $0.91$, and an F1 score of $0.94$. These findings demonstrate the algorithm's ability to choose appropriate labels.

To assess the agreement between the labels chosen by the algorithm and the expert labels, we calculated the inter-rater agreement. The analysis revealed a substantial agreement with an inter-rater agreement score of 0.89. This indicates a strong alignment between the labels assigned by the algorithm and the expert judgments.

\subsection{Discussion}
We have observed that many Muslims are eager to share their recitations. However, it is not the same as annotating the recitations.
There may be various reasons make Muslims hesitant to participate in annotating recitation, including:
\begin{itemize}
    \item Fear of making mistakes in labeling the noble verses.
    \item Lack of trust in the platform, as it is not supported by an official religious organization.
    \item Technological barriers that prevent some users from participating.
    \item Lack of perceived direct value or benefit for them.
\end{itemize}

Based on our observations in \ref{subsec-annoatated-dataset}, we found an inter-rater agreement of Krippendorff’s alpha = 0.63. This suggests that not all users may have fully grasped the instructions. The complexity of the task, with multiple labels to consider, could have contributed to this outcome. A solution for that could be to divide this annotation process task into two stages, one for checking audio validity and the other for checking audio correctness. Another potential solution is to crowdsource the data through learning Quran applications that provide value to users beyond just crowdsourcing. By incorporating the crowdsourcing component into a larger and more useful application, users may be more inclined to participate and provide accurate data. This approach could also help alleviate concerns around trust in the platform, as users may be more likely to trust a platform they are already familiar with and find useful.

\section{Conclusion}
\label{sec:con}
Learning to recite the Quran can be challenging for many non-Arabic Muslims around the world, especially those who do not have access to a tutor. Fortunately, AI technology can help simplify this process, but it requires a significant amount of carefully annotated Quranic data. In our work, we collected around 7,000 Quranic recitations from a pool of 1287 participants from more than 11 different non-Arabic countries. Additionally, We collected labels for 1166 recitations, with an inter-rater agreement of 0.63 and crowd accuracy of 0.77.
The evaluation of the used algorithm yielded impressive results, with an accuracy of 0.91, and an inter-rater agreement of of 0.89.

In future work, we aim to improve both inter-rater agreement and crowd accuracy. We will also incorporate different levels to identify and classify the exact mistakes made by the reciters including Tajweed mistakes.

\end{document}